# Comment on "Counter-propagating charge transport in the quantum Hall effect regime"

Yigal Meir[*,1] and Ady Stern[2]

[1]Department of Physics, Ben Gurion University of the Negev, Beer Sheva 84105, Israel
[2]Department of Condensed Matter Physics, The Weizmann Institute of Science, Rehovot 76100, Israel
[*]Corresponding author. Email: ymeir@bgu.ac.il

## Abstract

Laffont *et al.* (Science 363, 54–57 (2019)) report an upstream current in the $\nu=2/3$ quantum Hall regime, which, they claim, might be due to a counter-propagating charge mode.. We show that this observation can also be explained by the expected upstream neutral mode, without the need for an upstream charge mode. Our results agree with the observed data and explain why the upstream current is observed only in the unpolarized $\nu=2/3$ regime.

In a recent elegant experiment, Laffont *et al.* *(1)* have demonstrated an upstream current in a two quantum-point-contact (QPC) setup (Fig. 1) in the fractional $\nu=2/3$ quantum Hall regime. A current was injected at source S3, and detected at drain D3, indicating, the authors argue, an upstream charge mode. Indeed, such a mode was predicted *(2)* to exist on short distances (before the two counter-propagating modes equilibrate *(3,4)*) in the polarized $\nu=2/3$ regime. However, contrary to the theoretical predictions, the experiment observed such upstream current only in the unpolarized $\nu=2/3$ regime and not in the polarized one. Here we show that in the unpolarized regime, one may get an upstream current at drain D3, due to the upstream neutral model, without an upstream charge mode (Fig. 1). This is due to the fact that in the unpolarized regime, there are two propagating spin directions, which, due to the large magnetic field, have different tunneling amplitudes through the QPCs.

The unpolarized $\nu=2/3$ state is characterized by a spinless downstream charge mode, and a spinfull upstream neutral mode. In the experiment the source S3 injects electrons into the charge mode, consisting of an equal weight of spin up and spin down electrons. Due to the magnetic field, when this charge mode impinges upon QPC1, the transmission probabilities of spin up ($T_{1\uparrow}$) and spin down ($T_{1\downarrow}$) are different, leading to unequal weight of spin up and spin down electrons on the other side of QPC1. This necessarily results in an excitation of an upstream neutral mode, which propagates towards QPC2. There again, due to asymmetry of the transmission coefficients ($T_{2\downarrow}$ and $T_{2\uparrow}$), this neutral mode generates a downstream charge mode at QPC2, which can be collected at D3. On the other hand, in the polarized $\nu=2/3$ regime, where the two components of the charge and neutral modes have the same spin, there is no intrinsic asymmetry between their tunneling probabilities, and thus upstream charge will not be generated, consistent with the experimental observations.

To be more quantitative we notice that the magnitude of the upstream neutral mode generated at QPC1 is zero when $T_{1\uparrow} = T_{1\downarrow}$, and is maximal when one of them is equal to zero. So the simplest assumption would be that its magnitude would be equal to $(T_{1\uparrow} - T_{1\downarrow})^2$. Thus the magnitude of the charge current generated at QPC2 will be $(T_{1\uparrow} - T_{1\downarrow})^2 (T_{2\uparrow} - T_{2\downarrow})^2$. We define $\varepsilon_{i\uparrow}$ and $\varepsilon_{i\downarrow}$ as the onset energies where $T_{i\uparrow}$ and $T_{i\downarrow}$ rise from zero to unity, where the rise takes place on energy scales $\Delta_1$ and $\Delta_2$, respectively. As the energies of the two spin directions are split by the Zeeman effect, one expects $\varepsilon_{i\uparrow} - \varepsilon_{i\downarrow} \propto B$, where $B$ is the magnetic field. In the following we define the unit of energy as the Zeeman energy splitting when $B$ is equal to 1 Tesla. Given the experimentally measurable total transmission coefficient through each QPC, $T_i = (T_{i\uparrow} + T_{i\downarrow})/2$, one can then obtain $T_{i\uparrow}$ and $T_{i\downarrow}$ separately, and, from the above formula, the upstream current. Fig. 2 shows a comparison between the above calculation and the experimental measurements of the upstream current as a function of $T_1$ and $T_2$. To best fit the experiment, we set $\Delta_1, \Delta_2$=2 and 5, respectively. We see that there is a good agreement between theory and experiment. If either $T_1$ or $T_2$ is close to zero or to unity, then the transmission coefficients of both spin modes are similar and the upstream current is suppressed, reaching a maximum when both transmissions coefficients are close to ½, where the asymmetry between the transmission coefficients of the two spin is maximal.

To summarize, we have demonstrated that the experimental results can be explained within the current understanding of the *v*=2/3 fractional quantum Hall effect, without needing to invoke an upstream charge mode. The generation of a charge current from a neutral mode is a general feature when time reversal symmetry is broken, and is a special case of a "quantum Nernst engine" *(5)*. Our theory not only explains the observed upstream current in the unpolarized regime, but also explains why such current is not observed in the polarized regime, where the two modes have the same spin, and thus the same tunneling amplitudes. We predict that the upstream current in the unpolarized regime will increase when the Zeeman splitting increases, either by applying an additional parallel magnetic field, or by using samples where the transition between the unpolarized and polarized phases occurs at higher magnetic fields.

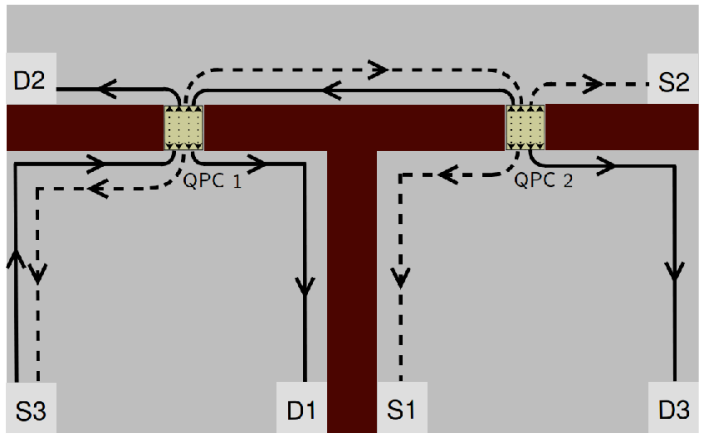

Fig. 1: **Upstream current without upstream charge mode.** In the experiment [1] a current was injected at S3 and detected at D3. Solid lines denote downstream charge modes, and broken line upstream neutral modes, so there is no upstream charge mode between the quantum point contacts. Nevertheless, the upstream neutral mode generated at QPC1 impinges on QPC2 and generates a downstream charge mode on the other side of the QPC, that is collected at D3 and gives rise to the observed experimental signal.

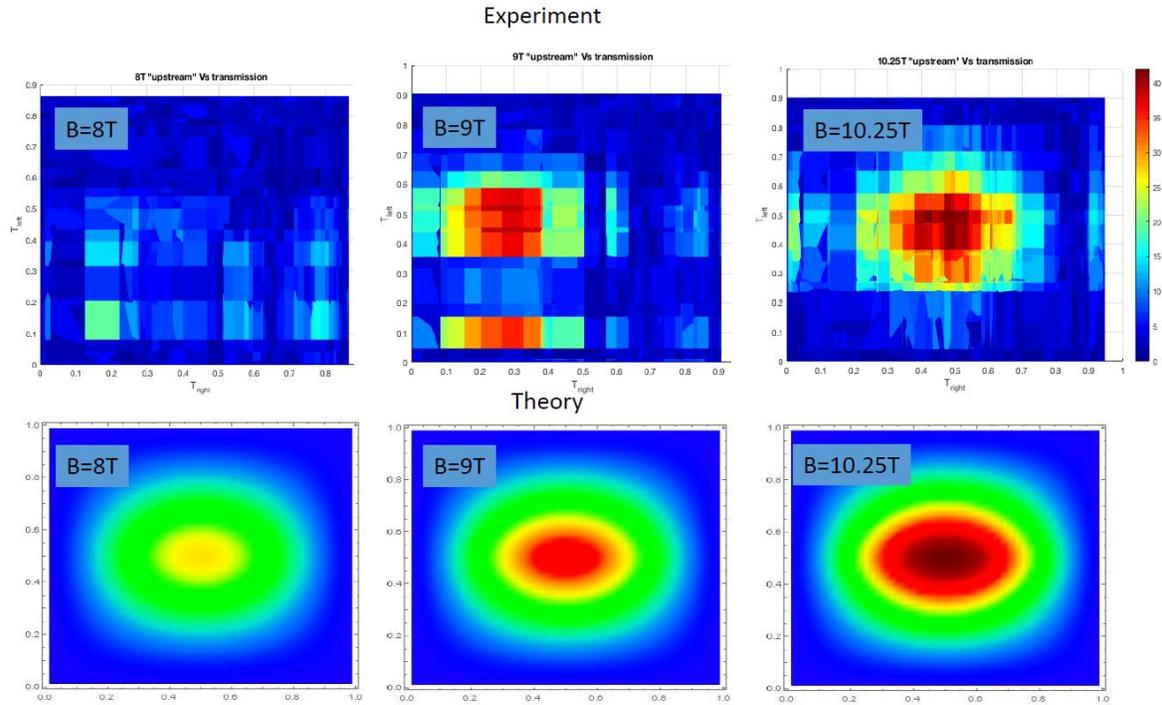

Fig. 2: **Comparison between experiment and theory.** The upstream current, as a function of the transmission coefficients of the two quantum point contacts, for 3 values of magnetic field as measured in the experiment (first row, from Fig. 4 in *(1)*) and as calculated theoretically using the formula in the text (second row).